\documentclass[preprint2]{aastex}
\usepackage{graphicx}
\usepackage{hyperref}
\slugcomment{Accepted for publication in AJ}

\shorttitle{Variable Sources in the LMC}
\shortauthors{Vijh et al.}

\begin{document}

\title{Variable Evolved Stars and YSOs Discovered in the Large Magellanic Cloud using the SAGE Survey}

\author{Uma P. Vijh\altaffilmark{1,2}, M. Meixner\altaffilmark{1}, B. Babler \altaffilmark{5}, M. Block\altaffilmark{8},  S. Bracker \altaffilmark{5},  C. W. Engelbracht\altaffilmark{8}, B. For\altaffilmark{8}, K. Gordon\altaffilmark{1},  J. Hora\altaffilmark{7},  R. Indebetouw\altaffilmark{6},  C. Leitherer\altaffilmark{1}, M. Meade \altaffilmark{5}, K. Misselt\altaffilmark{8}, M. Sewilo\altaffilmark{1}, S. Srinivasan\altaffilmark{3}, B. Whitney\altaffilmark{4}}

\altaffiltext{1}{Space Telescope Science Institute, Baltimore, MD 21218}
\altaffiltext{2}{Present Address: Rittter Astrophysical Research Center, University of Toledo, Toledo, OH 43606}
\altaffiltext{3}{Johns Hopkins University, Baltimore, MD 21218}
\altaffiltext{4}{Space Science Institute, Madison, WI 53716 }
\altaffiltext{5}{University of Wisconsin, Madison, WI 53706}
\altaffiltext{6}{University of Virginia, Charlottesville, VA 22903}
\altaffiltext{7}{CfA/Harvard, Cambridge, MA 02138}
\altaffiltext{8}{University of Arizona, Tucson, AZ 85719}

\begin{abstract}
We present initial results and source lists of variable sources in the Large Magellanic Cloud (LMC) for which we detect thermal infrared variability from the SAGE (Surveying the Agents of a Galaxy's Evolution) survey, which had 2 epochs of photometry separated by three  months. The SAGE survey mapped a $7\degr \times 7\degr$ region of the LMC using the IRAC and the MIPS instruments on board Spitzer. Variable sources are identified using a combination of the IRAC 3.6, 4.5, 5.8, 8.0~\micron\ bands and the MIPS 24~\micron\ bands. An error-weighted flux difference between the two epochs is used to assess the variability.  Of the $\sim 3$~million sources detected at both epochs we find $\sim 2,000$ variable sources for which we provide electronic catalogs. Most of the variable sources can be classified as  asymptotic giant branch (AGB) stars. A large fraction ($> 66\%$) of the extreme AGB stars are variable and only smaller fractions of carbon-rich (6.1\%) and oxygen-rich (2.0\%) stars are detected as variable. We also detect a population of variable young stellar object candidates.
\end{abstract}


\section{Introduction}
The Spitzer survey, Surveying the Agents of a Galaxy's Evolution (SAGE) of the Large Magellanic Cloud (LMC)  provides an unprecedented opportunity to detect thermal infrared variability of the infrared stellar population of the LMC.  Optical variability studies of the LMC by the MACHO \citep{alcock96} and OGLE \citep{paczynski94} monitoring projects  have revealed the period and luminosity relations for a wide variety of variable stars in the LMC,  e.g. the long period variables \citep{fraser05}. The three month time span between the two epochs of observations for  the SAGE survey  can detect  and constrain the variability of long period variables in the evolved star population and young stellar object candidates.  Thermal infrared variability of such objects has been studied in the Galaxy with {\sl IRAS} and {\sl ISO}, and Spitzer's improved sensitivity enables such studies in nearby galaxies.  The ~10 month {\sl IRAS}  mission surveyed most portions of the sky at least twice a year, separated by months and thus also constrained variability  of the brighter infrared sources in a similar way that the SAGE data samples the LMC's stellar populations.    
  
Most of the sources detected as variables in the thermal infrared by IRAS or ISO are evolved stars in our Galaxy.  For example, \citet{harmon88} used the IRAS VAR\footnote{The IRAS VAR  index  is a probability that the source is variable  in the IRAS 12 and 25 micron photometry.   e.g. VAR $>$ 50  means larger than 50\%  chance that the source is variable.  The variations have to be correlated,  i.e. in the same direction.} index to determine the nature of the IRAS stellar population that delineates  the Galactic Bulge.  In a statistical way,  they simulated the periods that the IRAS data were sensitive to and found the data were most sensitive to long period variables,  P$>$ 400 days  up to  1,400 days.  Essentially, large amplitude and long period variations are easier to detect and are also associated with larger mass-loss rates.  They used this period to estimate the age and mass range for this evolved stellar population in the bulge.  Later work by \citet{vanderveen90} also used the IRAS VAR to select and study OH/IR stars for ground based measurements of infrared light curves and find that these IRAS variables  fit well into the period-luminosity relations found for Mira variables in  Baade windows,  confirming that OH/IR stars  are more evolved forms of Miras.  ISOGal-DENIS studies of the Galactic bulge included one field with multiple ISOCam images separated by 17 -- 23 months  \citep{omont99}.  This study revealed  a population of  red giants with weak mass loss for which the brightest, and most dusty  red giants  have variation at 7 and 15~\micron.  These sources are ``intermediate aymptotic giant branch (AGB) stars''  i.e.  between early AGB and thermally pulsing AGBs,  and have low mass loss rates,  $10^{-10}$  to $10^{-11}$ M$_\odot$ yr$^{-1}$.  

In the LMC,  the evolved stars are too faint for IRAS and ISO measurements  to reliably detect variability.  However,  IRAS- and ISO- detected evolved stars have been followed up by ground based near-IR variability monitoring programs.  \citet{whitelock03} monitored  the IRAS-selected sample of AGB stars studied by ISO and find that larger amplitude variations in K are associated with redder stars.  The IRAS and ISO photometry are sensitive measures of dusty mass-loss of the AGB stars. Luminosity is correlated with period and longer periods have higher amplitude variations.  Thus, the thermal infrared variability of the evolved stars can cause a systematic scatter in the mass-loss rate determinations of these evolved stars in the LMC  and in mass-loss rate vs. luminosity relations \citep{groenewegen07}.

In contrast to the evolved star populations,  there have been few thermal infrared variability measurements of young stellar objects (YSOs).  IRAS measurements of variability of YSOs were at best difficult due to the low spatial resolution of IRAS and the confusion with diffuse ISM emission.  Nevertheless,  for isolated  YSOs  there was a detection of variability at 12 and 25~\micron\ for the Herbig Ae/Be stars  AB Aur and WW Vul  \citep{prusti94}.  Intermediate mass pre-main sequence stars like UX Ori stars show large photometric and polarimetric variations \citep{natta97}.   A concurrent ISO and ground based optical monitoring program of UX Ori type stars, SV Cep, was performed by \citet{juhasz07}.  \citet{juhasz07} found that the mid-infrared flux variations were anti-correlated with the optical variations, but that the far-infrared flux variations were positively correlated with the optical variability.  They used the infrared variability to help discriminate between disk or disk+envelope models and find that disk models are in better agreement with the data.  \citet{abraham04}  used ISO to  monitor the long term IR evolution of seven FU Orionis stars. They detected variation in four sources, which provides tentative support for the \citet{hartmann85} models for accretion outbursts in FU Ori objects as opposed to precessing jets \citep{herbig03}.  Interestingly,  more luminous stars which appear to have similar characteristics as the FU Orionis stars  have no detected outbursts.  Chromospheric hot-spots or accretion disk variability can cause optical  variability in pre-main sequence stars like T~Tauri stars, but these types of objects have very low infrared excesses and are unlikely to be in our list. Our list would include earlier stages of YSOs that have more substantial infrared excesses. Thus, while difficult to detect,  thermal infrared variability  of YSOs will provide key insights into the structure of YSOs and the physical mechanisms of the star formation process. 

In this paper,  we present initial results and source lists of SAGE point sources for which we have detected thermal infrared variability between the epoch 1 and 2 photometry measurements.   In \S 2,  we outline the method of our approach to detect this variability.  In \S 3,  we discuss the results of the variable source selection and their identification of evolved stars and YSOs.  In \S 4,  we discuss implications of the results and summarize the main conclusions in \S 5.  
  
\section{Method: Source Selection}
SAGE is a Legacy project on the Spitzer Space telescope \citep{werner04}, which mapped a $7\degr \times 7\degr$ region of the Large Magellanic Cloud (LMC) using the IRAC camera in the [3.6], [4.5], [5.8], and [8.0] micron bands \citep{fazio04} and the MIPS camera in the [24], [70], and [160] micron filters \citep{rieke04}.  The scientific goals of the survey focus on the life cycle of baryonic matter,  as traced by dust emission, from its start in the interstellar medium (ISM), to the formation of new stars,  to the death of  these stars and the return of matter to the ISM. The survey was done over two epochs with a total observing time of 291 hrs with IRAC and 217 hrs with MIPS.  The details of the survey are described in \citet{meixner06}.  The SAGE photometric data have been separately extracted from the two epochs, that are separated by 3 months. The IRAC epoch 1 data were taken on July 15 -- 26, 2005, epoch 2 on Oct 26 -- Nov 2, 2005 and the MIPS epoch 1 data were taken on July 27-- Aug 3, 2005 and epoch 2 on  Nov 2 -- 9, 2005. The SAGE Epoch 1 point source catalogs\footnote{Catalogs that combine 2MASS JHK \citep{Skrutskie06}, IRAC and MIPS 24~\micron\ data are available at the Spitzer Science Center website. http://ssc.spitzer.caltech.edu/legacy/all.html} containing over 4 million sources  of IRAC  (SAGEcatalogIRACepoch1)  and over 40,000 sources of MIPS 24 micron (SAGEcatalogMIPS24epoch1)  have been merged with 2MASS JHK, and the Magellanic Clouds Photometric Survey \citep{zaritsky04}.  Analysis of the colors and magnitudes of this catalog has revealed three general types of LMC point sources:  stars without dust, dusty evolved stars and young stellar objects \citep{meixner06}.   Further classification of the  SAGE sources by \citet{blum06} separated out the dusty evolved star classes such as supergiants,  and asymptotic giant branch (AGB) stars from the red giants.  The young stellar object (YSO) population has been identified in part by \citet{whitney08}.
 
The SAGE 2-epoch point sources catalogs for IRAC (SAGEcatalogIRACepoch1 and SAGEcatalogIRACepoch2) and  MIPS (SAGEcatalogMIPS24epoch1 and SAGEcatalogMIPS24epoch2) were used for this study. Variable sources were identified using the matching and selection criteria described below and a variability index.  We define the variability index $V $ as the error-weighted flux difference in each SAGE band:
\begin{displaymath}
V =\frac{(f_1 - f_2)}{\sqrt{\sigma f_1^2+ \sigma f_2^2}}
\end{displaymath}
\noindent where $f_1$ and $f_2$ are fluxes in the two epochs and $\sigma f_1$ and $\sigma f_2$ are the associated errors.
We also defined a fractional flux (fF) for each band, the fractional change in flux
\begin{displaymath}
fF=\frac{f_1 - f_2}{(f_1 + f_2)/2}
\end{displaymath}
\noindent where $f_1$ and $f_2$ are fluxes in the two epochs.
The criteria for inclusion of sources into the catalog and the archive version are detailed in \citet{meixner06}. The source lists are stored in the SAGE database, a system implemented with Microsoft SQL server 2000. We developed scripts in Structured Query Language (SQL) for creating the inter-epoch matching sources and for subsequent checks on, and quality control of, the variable source lists.

\subsection{Inter-epoch Matching of Sources}
All four IRAC bands and the MIPS 24 micron, 2-epoch data were used for this study. To find the variable IRAC sources in SAGE, the IRAC epoch 1 and epoch 2 catalogs were matched using a 2\arcsec\  search radius. Sources with multiple matches and sources with any neighboring sources within 3\arcsec\ were excluded as careful consideration revealed that the PSF characteristics of the IRAC detectors make the measured fluxes of sources with neighbors closer than 2\arcsec\ less reliable.  Furthermore, to avoid possibility of a mismatch between the epochs, only matches within 0.9\arcsec\ (the 3-$\sigma$ value for IRAC photometry) were retained.   We also used the following magnitude cuts to only include sources with highly reliable fluxes in our lists: 16, 16, 14  and 13.5 mag at 3.6, 4.5, 5.8, and 8.0 \micron\ respectively.

\begin{figure}[!h]
\includegraphics[width=3in]{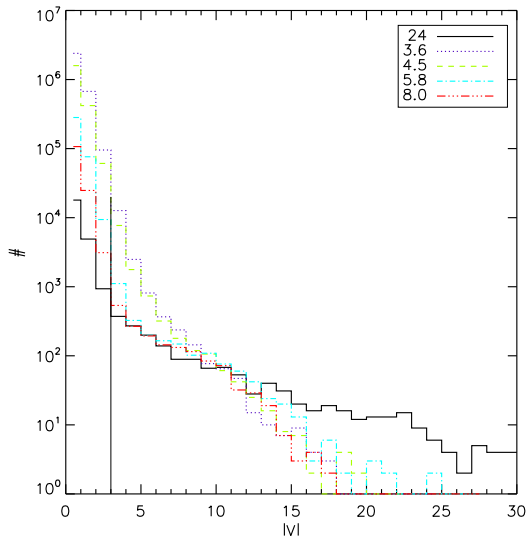}
\caption{Histogram of all the SAGE sources matched between 2 epochs for the different bands of IRAC and MIPS 24~\micron\ band. Sources with  $\vert V\vert >$ 3 in at least 2 consecutive bands are classified as variable sources.\label{Vhist}}
\end{figure}

For MIPS, the 24\micron\ PSCs SAGEcatalogMIPS24epoch1 and SAGEcatalogMIPS24epoch2 were used for this study.  We used a 1\arcsec\ matching between the 2 epochs to eliminate wrong inter-epoch matches. Both the individual epoch catalogs have their astrometry matched to IRAC catalogs, which have been matched to the 0.3\arcsec\ accuracy 2MASS catalog. We also note here that the MIPS24 catalogs are conservative catalogs that report only sources with reliable fluxes. With only two epochs of data we do not aim to produce a complete list of infrared variable sources in the LMC; nonetheless we aim for a highly reliable one, and require variability in more than one band.  Figure~\ref{Vhist} shows a histogram of the variability $\vert V\vert$ for the inter-epoch matched sources for all the IRAC and MIPS 24~\micron\ bands. $\vert V\vert <3 $ probably indicates ``random'' errors in flux and $\vert V\vert >3 $ a population being dominated by a systematic trend. Different bands are shown in different colors and the vertical lines at $\vert V\vert= 3$ indicate the reliability cuts for the SAGE variables.   

\subsection{Variability Criteria}
Our criterion for a variable source is $\vert V\vert >3$ in at least 2 consecutive bands in the same direction. The IRAC-selected sources were matched with the inter-epoch matched 24~\micron\ sources. 24~\micron\ sources with $\vert V_{24}\vert >3 $ as well as $\vert V_{8.0}\vert >3 $ in the same direction are deemed 24 \micron\ variables.  Table~\ref{t-crit} explains the criteria used to identify the variable sources. The first part of the table characterizes the source catalogs. For each band the number of sources with valid fluxes in the band at both epochs, the number and percentage of sources having $\vert V_{band}\vert > 3$, and the number and percentage of sources expected statistically to have $\vert V_{band}\vert >3$ are given. The second part describes the selection criteria and the statistical significance of the variable sources. Here for each band, the sources not only have valid fluxes in both epochs for a given band but also in a neighboring band. Once again the numbers and percentages of sources both meeting the variability criterion and those that might do so based purely on statistical considerations (and therefore likely to be false variables) are given. Assuming that the measurement of  flux for any source follows a gaussian distribution characterized by the error in the flux, the probability that two measurements of the flux of a non-variable source will not be within 3-sigma of each other is 0.27\%. This applies for every band. We see that for all bands the number of sources with $\vert V\vert > 3$ is more than that expected from statistical variations. Requiring that the sources have $\vert V\vert >$ 3 in at least two bands in the same direction considerably reduces the chances of random mis-classification as a variable source.  Firstly a source has to have valid fluxes at both epochs in at least two neighboring bands, and secondly have $\vert V\vert > 3$ in both of these bands in the same direction. Statistically this has a probability of 0.0004\%\ for bands with a single neighboring band and 0.0007\%\ for bands with two neighbors.  For example, at 24~\micron\ less than 1 source would be classified as a variable source due to random chance alone.  The argument could be made that the flux measurement of sources does not follow a gaussian distribution, but this simple, idealized statistical approach gives us lower limits on the number of false variables in the list. We perform other tests on the variable source list that do not rely on normal error distributions and  uncorrelated errors, and those tests suggest that most of the sources in the list are really variables. A few sources have band-to-band inconsistencies in $V$, that indicates wrong photometry in one band. However, in these cases the the inconsistent band is not responsible for the variability criterion and these fluxes have been nulled in the tables. Of the $\sim 4$ million sources in the SAGE Catalog, Table~\ref{t-prop} catalogs the properties (coordinates, fluxes and variability) of the 1,967 SAGE variables. 
\begin{deluxetable}{lccccc}
\tabletypesize{\small}
\tablecaption{Criteria for variable sources \label{t-crit}}
\tablewidth{0pt}
\tablehead{\colhead{Criteria} & \colhead{3.6\ \micron} &\colhead{4.5\ \micron} & \colhead{5.8\ \micron} & \colhead{8.0\ \micron} & \colhead{24\ \micron}}
\startdata
\# of sources with valid fluxes		&	2,962,938	&	1,939,173	&	354,290	&	133,655		&	25,412\\
\# with $\vert V_{band}\vert >$ 3	&	15341		&	9857		&	2395 		&	1647 		&	1625  \\
\% with $\vert V_{band}\vert >$ 3	&	0.52		&	0.51		&	0.68		&	1.23		&	6.39\\
\# expected statistically			& 	8000		&	5236		&	957		&	361		&	69 \\
\% expected statistically			&	0.27		&	0.27		&	0.27		&	0.27		&	0.27\\
\# of sources with valid fluxes 		&			&			&			&			&	\\
in neighboring bands				&1,787,038\tablenotemark{a}&2,121,027\tablenotemark{b}&458,607\tablenotemark{c}&124,618\tablenotemark{d}&13901\tablenotemark{e}\\
\# meet variability criterion			&	1520		&	1827		&	1582		&	1316		&	623\\
\% meet variability criterion		&	0.09		&	0.09		&	0.34		&	1.06		&	4.48\\
\# expected statistically			&	8		&	15		&	3		&	$<$1		&	$<$1\\
\% expected statistically			&	0.0004	&	0.0007		&	0.0007		&	0.0004	&	0.0004\\
\enddata
\tablenotetext{a}{Valid fluxes at 3.6 and 4.5\ \micron}
\tablenotetext{b}{Valid fluxes at 3.6 and 4.5\ \micron\ or at 4.5 and 5.8\ \micron}
\tablenotetext{c}{Valid fluxes at 4.5 and 5.8\ \micron\ or at 5.8 and 8.0\ \micron}
\tablenotetext{d}{Valid fluxes at 5.8 and 8.0\ \micron}
\tablenotetext{e}{Valid fluxes at 8.0 and 24\ \micron}
\end{deluxetable}
\begin{deluxetable}{lccccccccccccccccccccccccc}
\tablecolumns{26}
\tabletypesize{\scriptsize}
\rotate
\tablecaption{Properties of the 1967 SAGE Variables \label{t-prop}}
\tablewidth{0pt}
\tablehead{
\colhead{IRAC} & \colhead{MIPS} & \colhead{} & \colhead{R.A.} & \colhead{DECL.} & \colhead{} & \multicolumn{4}{c}{3.6 \micron} & \multicolumn{4}{c}{4.5 \micron} & \multicolumn{4}{c}{5.8 \micron} & \multicolumn{4}{c}{8.0 \micron} & \multicolumn{4}{c}{24 \micron} \\
\cline{7-10}  \cline{12-13}  \cline{15-17}  \cline{19-21}  \cline{23-26}\\
\colhead{Designation} & \colhead{Designation} & \colhead{E} & \colhead{(deg)} & \colhead{(deg)} & \colhead{Class} & \colhead{flux} & \colhead{err} & \colhead{ V} & \colhead{fF} & \multicolumn{16}{c}{... same columns ...}\\
\colhead{} & \colhead{} & \colhead{} & \colhead{} & \colhead{} & \colhead{} & \colhead{(mJy)} & \colhead{(mJy)} & \colhead{} & \colhead{} & \colhead{} & \colhead{}  & \colhead{} & \colhead{} & \colhead{} & \colhead{} & \colhead{} & \colhead{} & \colhead{} & \colhead{} & \colhead{} & \colhead{} & \colhead{} & \colhead{} & \colhead{} & \colhead{}
}
\startdata
SSTISAGE1C ... &  SSTM1SAGE1 ... & 1 & 79.9337 & -69.9940 & X & 108.80 & 5.08 & 4.1 & 0.25 & & & & & & & & & & & & & & & &\\
SSTISAGE2C ... &  SSTM1SAGE2 ... & 2 & 79.9335 & -69.9941 & & 85.05 & 2.71 & & & & & & & & & & & & & & & & & &\\
SSTISAGE1C ... &  SSTM1SAGE1 ... & 1 & 79.9421 & -68.8979 & Y & 0.72 & 0.04 & -4.4 &-0.29 & & & & & & & & & & & & & & & &\\
SSTISAGE2C ... &  SSTM1SAGE2 ... & 2 & 79.9422 & -68.8979 & & 0.96 & 0.04 & & & & & & & & & & & & & & & & & &\\
SSTISAGE1C ... &  ... & 1 & 79.9643 & -72.6359 & U & 4.93 & 0.18 & -5.4 & -0.29 & & & & & & & & & & & & & & & &\\
SSTISAGE2C ... &  ... & 2 & 79.9646 & -72.6359 & & 3.67 & 0.15 & & & & & & & & & & & & & & & & & &\\
SSTISAGE1C ... &  ... & 1 & 79.9718 & -69.6882 & O & 5.23 & 0.18 & 5.0 & 0.29 & & & & & & & & & & & & & & & &\\
SSTISAGE2C ... &  ... & 2 & 79.9719 & -69.6883 & & 6.97 & 0.30 & & & & & & & & & & & & & & & & & &\\
SSTISAGE1C ... &  ... & 1 & 79.9811 & -69.3925 & C & 25.27 & 0.63 & 3.3 & 0.10 & & & & & & & & & & & & & & & &\\
SSTISAGE2C ... &  ... & 2 & 79.9813 & -69.3925 & & 22.83 & 0.39 & & & & & & & & & & & & & & & & & &\\
\enddata
\tablecomments{Table \ref{t-prop} is published in its entirety in the electronic edition of the {\it Astronomical Journal}.  A portion is shown here for guidance regarding its form and content.}
\end{deluxetable}

\section{Source Classification}
Using the variability criteria discussed in \S 2 we find $1,967$ SAGE variables at IRAC and MIPS 24~\micron\ bands. 514 sources are variable ($\vert V\vert > 3$) in all the five bands. Table~\ref{t-prop} lists the properties of the 1,967 SAGE variables.

\begin{figure}[!h]
\includegraphics[width=3in]{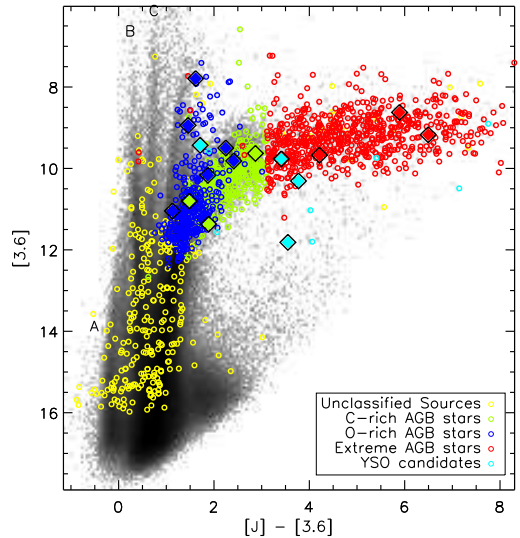}
\caption{Color magnitude diagrams (CMDs) of the sources detected as variables in the SAGE survey. The SAGE variables over-plotted on a Hess diagram showing the distribution of all the stars in the SAGE Epoch 1 catalog. AGB stars are classified into C-rich, O-rich and extreme AGB following the scheme followed by Cioni et al. 2006 and Blum et al. 2006. YSO candidates are from cross-correlation with the SAGE-YSOs classified by Whitney et al (2008).  SEDs of sources plotted with solid diamonds are shown in Figures~\ref{oagbsed}-\ref{xagbsed} \&\ref{ysosed}.\label{hessJ-3.6}}
\end{figure}

Most of the SAGE variables can be classified as AGB stars, a small number as YSOs and a number of sources that are presently unclassified but could be OB stars, RGB stars, post-AGB stars, PNe, background active galaxies or other classes of variables like cepheids,  RR Lyrae stars or WR stars. We choose three color-magnitude diagrams (CMDs) to classify the SAGE variable sources. In Figure~\ref{hessJ-3.6} all the sources in the SAGE epoch 1 catalog are used to plot the Hess diagram shown in grayscale. \citet{blum06} showed that the [J] - [3.6] vs. [3.6] CMD was most useful to see the separation of the O-rich, C-rich and extreme-AGB population.  Over-plotted are the epoch~1 magnitudes of the SAGE variable sources classified into O-rich, C-rich and extreme AGBs based on the classification scheme of \citet{cioni06} and \citet{blum06}. The features in the underlying CMD (grayscale) are labeled `A', `B', and `C' \citep[see][]{blum06,nikolaev00}: `A' being the OB star locus at the bluest, faint end of the diagram. The first prominent finger (labeled `B') corresponds to young A-G supergiants (SGs). The finger `C' consists mainly of foreground dwarfs and giants. The next finger to the right in the figure represents late-type(mostly M) SGs and luminous, O-rich M stars \citep{blum06}. The rest of the region above the tip of the RGB are divided into O-rich, C-rich and extreme AGB zones. Keeping these divisions in mind lets us classify most of the variable sources as evolved stars. To identify the YSO population, we cross-correlate the variable sources with the SAGE-YSO list from \citet{whitney08} and find 29 YSO candidates. Sources that are not classified as evolved stars or YSO candidates are plotted in yellow and will be discussed further in \S~\ref{unc}.  

\begin{figure}[!h]
\includegraphics[width=3in]{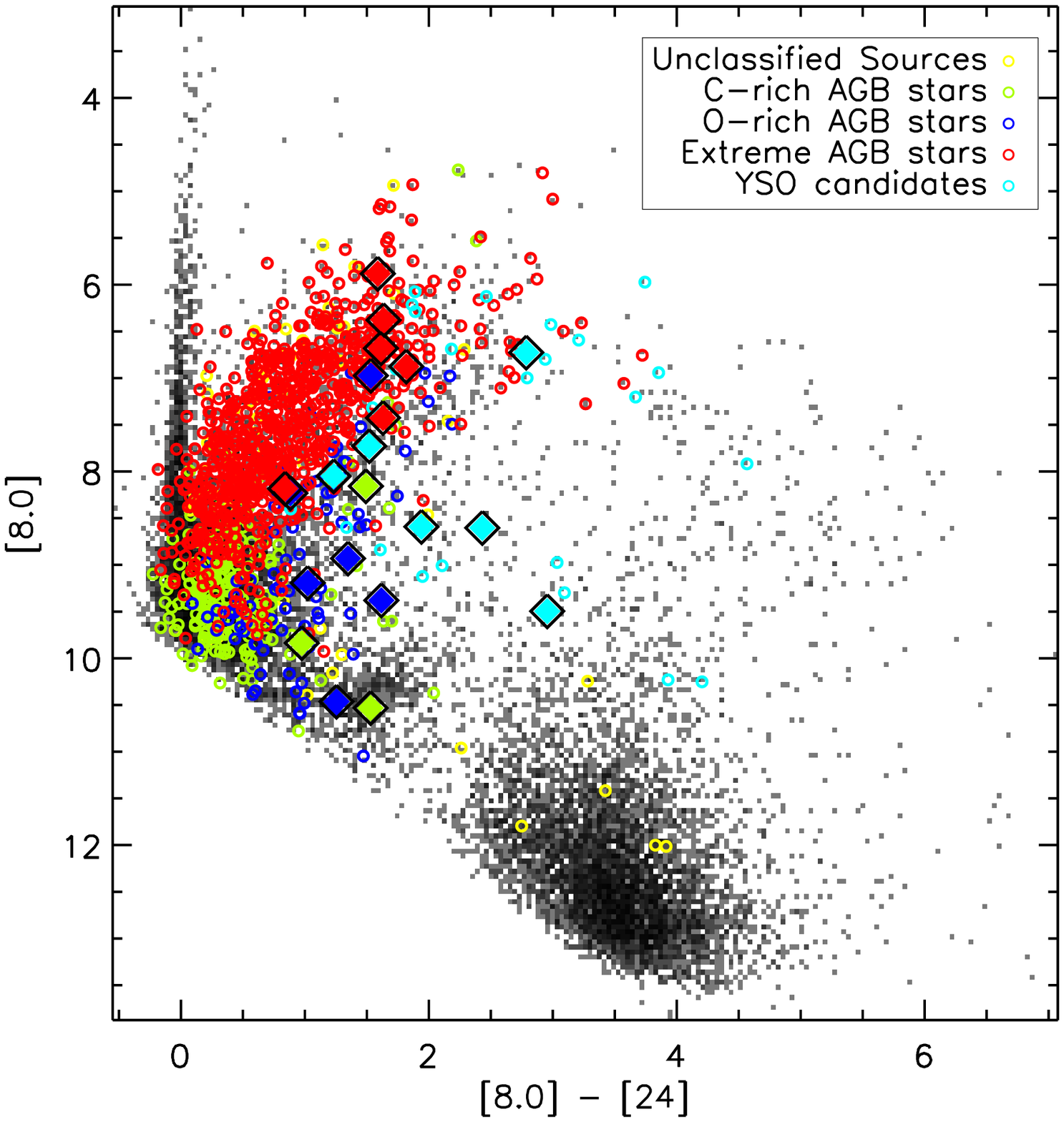}
\caption{CMD of the SAGE variable sources. Classification of the AGB stars and underlying Hess diagram are same as that in the previous figure.  SEDs of sources plotted with solid diamonds are shown in Figures~\ref{oagbsed}-\ref{xagbsed} \&\ref{ysosed}.\label{hess8-24}}
\end{figure}

\begin{figure}[!h]
\includegraphics[width=3in]{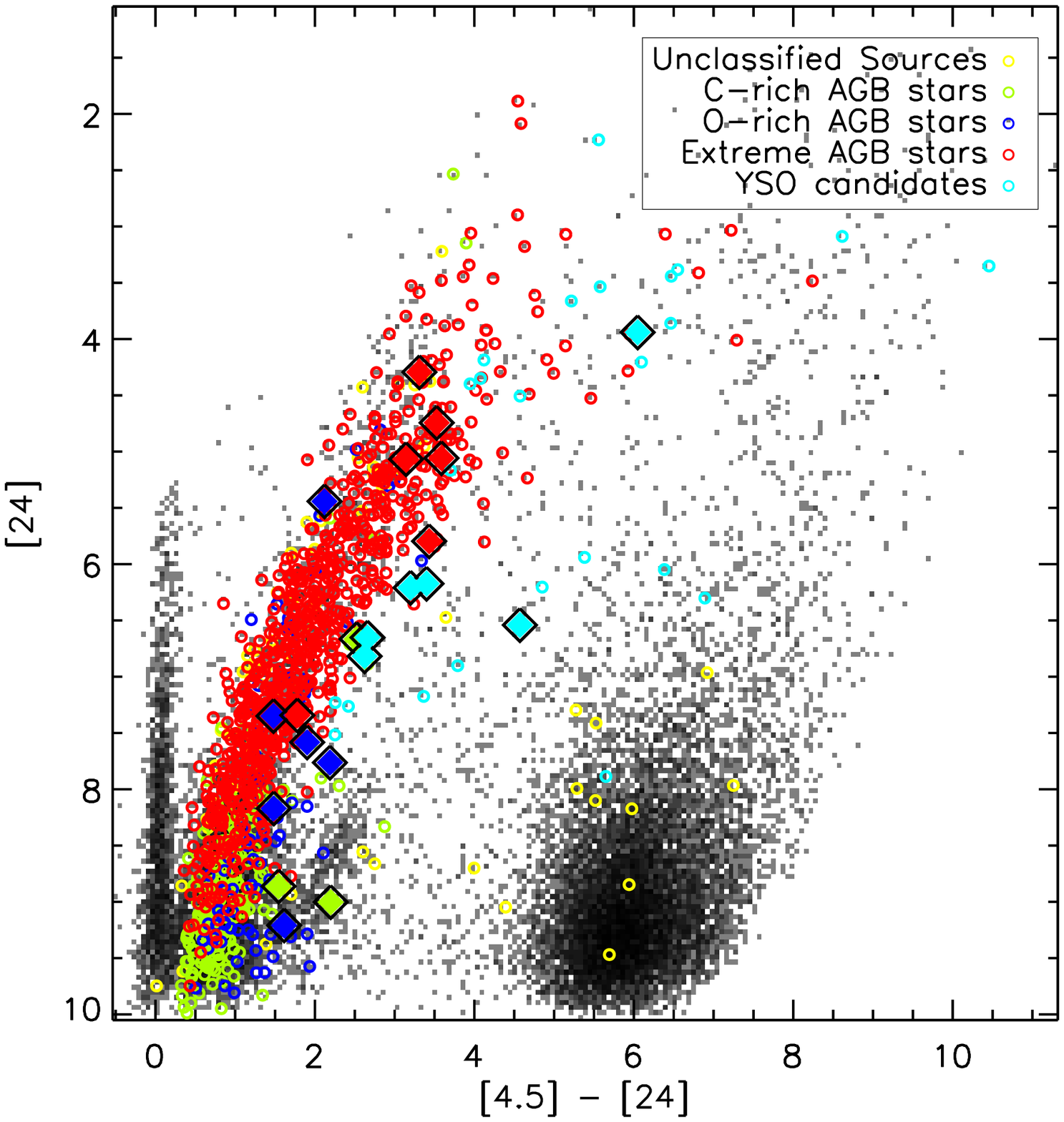}
\caption{CMD of the SAGE variable sources. Classification of the AGB stars and underlying Hess diagram are same as that in the previous figure.  SEDs of sources plotted with solid diamonds are shown in Figures~\ref{oagbsed}-\ref{xagbsed} \&\ref{ysosed}.\label{hess4.5-24}}
\end{figure}

Meixner et al. (2006) showed that the [8.0] -[24] vs [8.0] CMD was most appropriate for separating the dusty objects based on mass loss. Figure~\ref{hess8-24} shows the [8.0] - [24] vs [8.0] CMD for the SAGE variables, with the same color-scheme used for the classification as in Figure~\ref{hessJ-3.6}. Figure~\ref{hess4.5-24} shows the [4.5] - [24] vs [24] CMD for the SAGE variables. This CMD is the one of the most useful ones to delineate the LMC stellar population from the background population. We see that most of the unclassified sources are among the redder and fainter population in this CMD and are likely to be external galaxies. These sources are further discussed in the next section \S~\ref{unc}.

\subsection{Unclassified Sources\label{unc}}
Using 2MASS and 3.6~\micron\  color criteria most of the variable sources are classified as evolved AGB stars (with O-rich, C-rich and extreme subclasses). Using the \citet{whitney08} YSO candidate list we also identify YSO candidates as infrared variables. The classification of the variable sources that do not fall under any of these subcategories is more challenging.  Approximately 17\%  of the variables remain unclassified and $\sim~70\%$ of them remain so mainly because they are missing IRAC and/or 2MASS fluxes. This could be because a) they are undetected being dust enshrouded, e.g very dusty YSOs, b) extremely bright and therefore saturated in IRAC, or c) spatially extended at IRAC wavelengths and thus missing from the point source catalog. In preliminary follow-up work we have identified 20 unclassified sources as cepheids detected in the MACHO survey. Some of the other `unclassified' variables fall on distinct regions of the CMD and could be variable OB stars, RGB stars or O-rich stars that were too faint to be detected in the 2MASS survey. The other sources could be YSOs, Post-AGB sources, PNe,  background galaxies or other classes of variables like cepheids, RRLyrae stars, WR stars, QSOs or AGN. Further follow-up work is be required to identify the specific nature of these sources.

\section{Discussion} 
\begin{figure}[!h]
\includegraphics[width=3in]{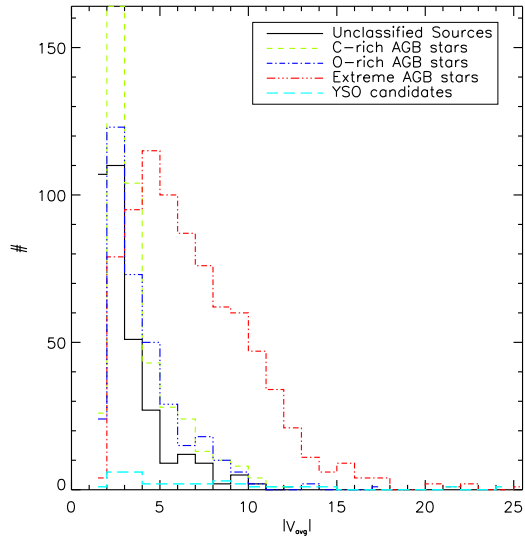}
\caption{Histogram showing the variability distribution among the different classes of SAGE IRAC variables. Average variability in all 4 bands of IRAC and MIPS 24~\micron\ band has been plotted. C-rich stars in green, O-rich stars in blue, extreme AGBs in red, YSO candidates in cyan and the unclassified sources in black.\label{VChist}}
\end{figure}

In this first paper we identify infrared variable sources in the LMC and provide an approximate classification as to their nature as discussed in \S3.  In this section, we discuss the nature of these sources and some implications of their infrared variability for SAGE studies.  In particular, we focus on the evolved AGB  star population and YSO candidates.  The total number of variables identified as SAGE variables  is 1,967 or $\sim$0.06\% of the $\sim$3 million IRAC point sources that are common to both epoch 1 and 2 catalogs .  These fractional numbers suggest that we preferentially detect as variable the  redder or dustier sources which tend to be AGB stars or YSOs.   In Table \ref{interptab} we summarize the different categories of  sources we have detected as variables and compare them to other work done on the classification of SAGE sources as evolved stars \citep{blum06} and YSO candidates \citep{whitney08} and discuss them below. Histograms (Figure~\ref{VChist}) of average variability over all bands show how the AGB stars, YSOs and unclassified sources differ in their variability.
\begin{deluxetable}{lccc}
\tablecaption{SAGE Variable Population \label{interptab}}
\tablewidth{0pt}
\tablehead{
\colhead{Source Type} & \colhead{SAGE Variables} & \colhead{SAGE Epoch 1 Sources} & \colhead{\% detected as variable}
}
\startdata
Total		&	1967		&	4,338,548		&	0.05\\
O-rich AGB	&	353		&	17,875\tablenotemark{a}	&	1.975 \\
C-rich AGB	&	426		&	6,935\tablenotemark{a}	& 	6.14\\
Extreme AGB 	&	820		&	1,240\tablenotemark{a}	&	66.1\\
YSO candidate	&	29		&	990\tablenotemark{b}	&	2.93 \\
Unclassified	&	335		&	4,311,508 		&	0.008\\ 
\enddata
\tablenotetext{a}{Classified by Blum et al. (2006)}
\tablenotetext{b}{Classified by Whitney et al. (2008)}
\end{deluxetable}

\subsection{Evolved stars}
\begin{figure}[!h]
\includegraphics[width=3in]{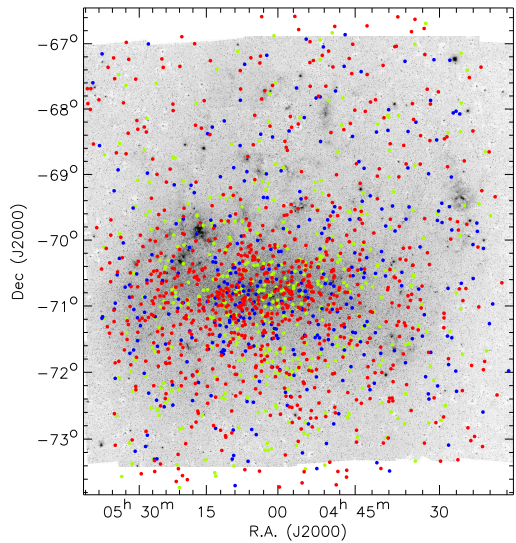}
\caption{The spatial distribution of the variable sources in the LMC classified as AGB stars plotted on the 3.6 \micron\ map. Note the correlation of the spatial distribution of the variable population with the stellar density function (3.6 \micron). This confirms that most of these sources are indeed stars. The colors represent the same classification of the sources as used in Figure~\ref{hessJ-3.6}. \label{spatirac}}
\end{figure}

As expected,  the variable source population is dominated by evolved stars, in particular the AGB stars.  The SAGE variables  have $>81$\% AGB stars, classified as either O-rich, C-rich or extreme AGB stars.  The spatial distribution of the SAGE variables, as shown in Figure~\ref{spatirac}, correlates approximately with the LMC 3.6~\micron\ image which traces predominantly  the old stellar population of the LMC \citep{blum06} and  supports our identification of the variables as  evolved stars.   In particular, the highest density of sources is located in the main stellar bar of the LMC.  The color coding of the point sources on this image follows that of the CMDs.  The extreme AGB stars, shown in red, dominate both the spatial distribution of  variable sources and the CMDs (Figures~\ref{hessJ-3.6}  -- \ref{hess4.5-24}).  Looking at the numbers of sources detected per category (Table \ref{interptab}) we find that the extreme AGB stars outnumber the O-rich or C-rich AGB stars  by a factor of two.  Moreover,  the percentage of variables  changes dramatically across the classified AGB sources.  Only 2.0\% of the O-rich AGB stars have been detected as variable.  This percentage increases to 6.1\% for the C-rich AGB stars.  For the extreme AGB stars, the percentage jumps  to 66\%.  This increase in percentage follows an evolutionary trend  of the AGB.  All stars on the early AGB start off as O-rich.  As they evolve to higher luminosities their pulsational periods increase, and many become C-rich AGB stars.  At the highest luminosities, the periods are the longest and these extreme AGB stars are characterized by significant circumstellar dust emission.  Thus,  the fractional number of AGB stars identified as variable increases as the star becomes more evolved on the AGB.   
        
This preferential detection of the evolved  AGB stars is similar to that in the IRAS variability studies in our Galaxy because our sampling period of  2 measurements separated by 3 months is similar to the IRAS sampling.   As \citet{harmon88} found in our Galaxy,  we are preferentially finding the more evolved, dusty  AGB stars which probably have longer periods.  This preference is at least in part due to the sensitivities of detection and of time sampling.   Because they are more luminous  and  dust enshrouded,  the extreme AGB stars are easier to detect and their variability is easier to measure reliably in the IRAC and MIPS bands.  In addition, the period of the variability for the extreme AGB stars, $\sim$400-500 days, is well matched to our two sample points in time separated by 3 months ($\sim$90 days) such that we could more likely than not measure a detectable flux difference in the two epochs.    

\begin{figure}[!h]
\includegraphics[width=3in]{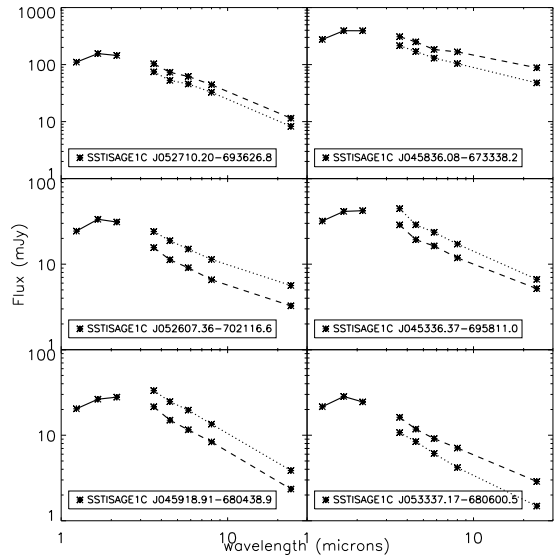}
\caption{Spectral energy distributions showing the two epochs of a sample of variable O-rich AGB stars. SAGE epoch 1 fluxes are connected by dotted lines and epoch 2 by dashed lines. The sources are marked with solid, blue diamonds on the CMDs. The 2MASS data are connected by solid lines and in many instances appears disjoint from either of the epochs, which is due to infrared variability as these data represent a different phase far removed from either of the SAGE epochs.  \label{oagbsed}}
\end{figure}

\begin{figure}[!h]
\includegraphics[width=3in]{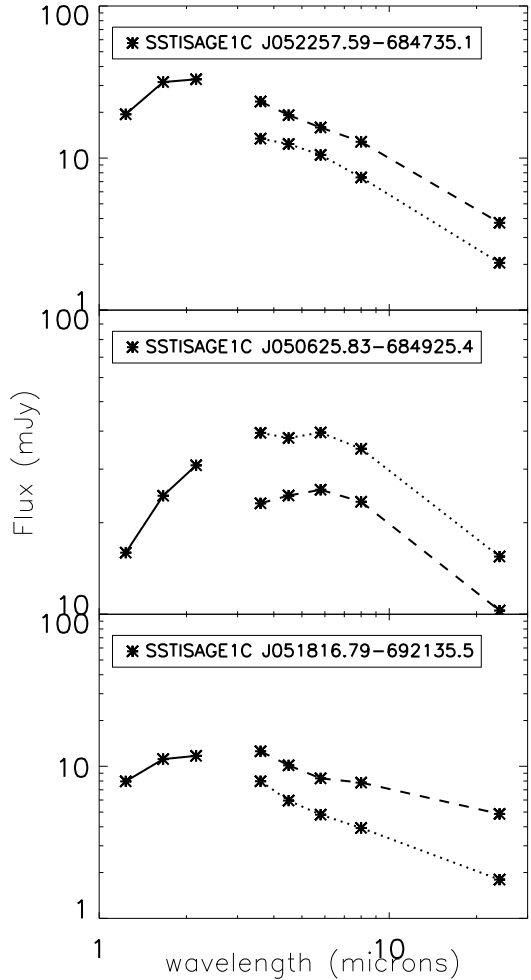}
\caption{Spectral energy distributions showing the two epochs  of a sample of variable C-rich AGB stars. The sources are marked as solid, green diamonds on the CMDs. SAGE epoch 1 fluxes are connected by dotted lines and epoch 2 by dashed lines. The sources are marked with solid, red diamonds on the CMDs. The 2MASS data are connected by solid lines and in many instances appears disjoint from either of the epochs, which is due to infrared variability as these data represent a different phase far removed from either of the SAGE epochs. \label{cagbsed}}
\end{figure}

\begin{figure}[!h]
\includegraphics[width=3in]{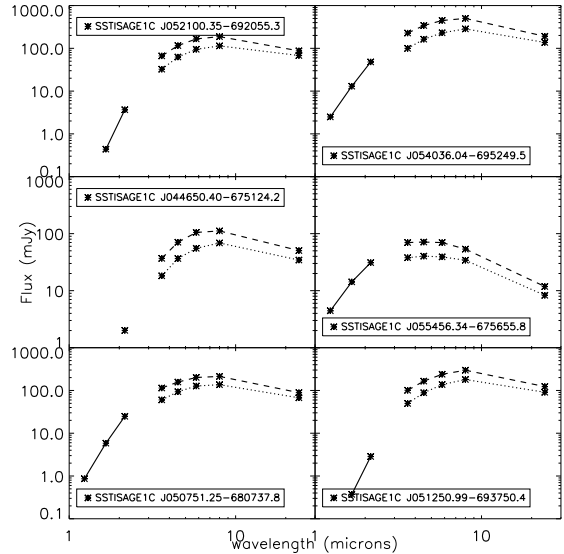}
\caption{Spectral energy distributions showing the two epochs  of a sample of variable extreme AGB stars. SAGE epoch 1 fluxes are connected by dotted lines and epoch 2 by dashed lines. The sources are marked with solid, red diamonds on the CMDs. The 2MASS data are connected by solid lines and in many instances appears disjoint from either of the epochs, which is due to infrared variability as these data represent a different phase far removed from either of the SAGE epochs. \label{xagbsed}}
\end{figure}

Even though we do not detect all the AGB stars as infrared variables, we do know that they are all variables \citep[e.g.][]{fraser05}.  The detected infrared variability reported here provides a constraint on the possible range of the infrared flux in these types of  AGB stars.   The fractional difference in fluxes we detect  has a range of 0.1 to 1.5 of the total flux.   These changes in infrared flux  will have significant implications for the analysis of SAGE data on the AGB stars.  Estimates of the infrared excesses for this population will have some systematic errors because of the infrared variability.    The mass-loss rates of the evolved stars will be derived from model fits to the spectral energy distributions of these sources, and the IRAC and MIPS 24 \micron\ photometry provide constraints on the dusty shells.  The infrared variability of the three different classes of AGB stars is revealed  in the 2-epoch spectral energy distributions of a few O-rich, C-rich and extreme AGB stars selected from our sample (Figures~\ref{oagbsed} -- \ref{xagbsed}).    The SEDs show the SAGE epoch1 data connected by dotted lines and epoch2 data by dashed lines. The 2MASS data are connected by solid lines and in many instances appears disjoint from either of the epochs, which is due to infrared variability as these data represent a different phase far removed from either of the SAGE epochs. The O-rich and C-rich AGB stars show a similar amount of variation and the extreme AGB star shows a larger variation.  Indeed the extreme AGB stars typically have a larger variability index than the O-rich and C-rich AGB stars (Figure \ref{VChist}).

Some of the brightest  variables are not classified because they are saturated in IRAC 3.6~\micron\  band or 2MASS bands,  which we use for classification purposes.   The drop in flux in the Raleigh-Jeans tails of the SED enables them to be detected at the longer IRAC bands  (4.5, 5.8 and 8.0~\micron) and MIPS  24~\micron\ without saturation. 
 
\subsection{YSO candidates}
\begin{figure}[!h]
\includegraphics[width=3in]{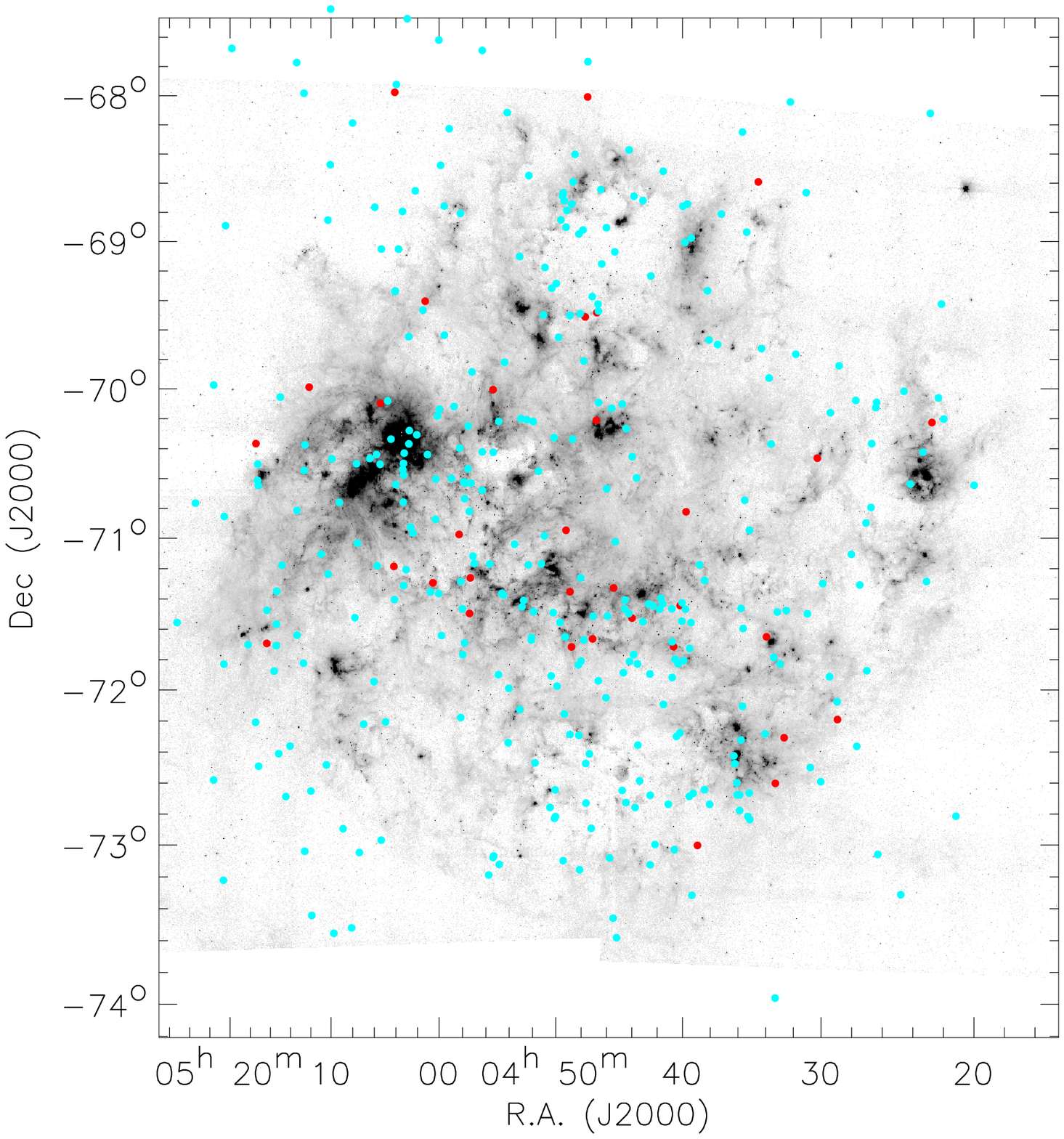}
\caption{The spatial distribution of variable massive YSOs and the unclassified sources plotted on SAGE 24 \micron\ image. The YSO candidates are shown in red and the unclassified candidates in cyan. \label{spatMipsyso}}
\end{figure}

We have discovered a YSO candidate population with infrared variability. That this phenomenon was not noticed earlier is not surprising given that little work has been done on infrared variability of YSOs  and most of it has been followup of known optically variable sources. These variable YSO candidates comprise at least 3\% of all the YSO candidates.  We say at least,  because there are a large number of sources in the same CMD space that are ``unclassified''.  The spatial distribution of YSO candidates correlates spatially with  the  LMC MIPS 24 micron emission, which traces massive star formation, supporting the identification of these variables as YSOs (Figure~\ref{spatMipsyso}).  Also interesting is the spatial distribution of the ``unclassified'' sources (Figure~\ref{spatMipsyso}). These sources also closely follow the the 24~\micron\ emission. Further follow up work on extending the list of YSO variables that is in progress will help clarify if some of these unclassified sources are YSOs.

\begin{figure}[!h]
\includegraphics[width=3in]{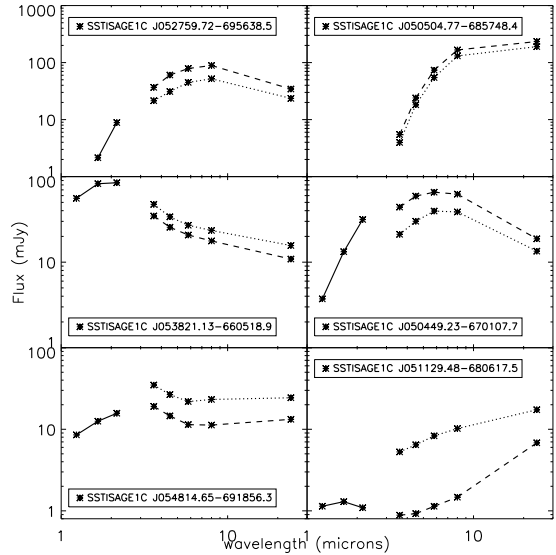}
\caption{Spectral energy distributions (SEDs) of a sample of YSO variable sources. The sources are marked as solid, cyan diamonds on the CMDs. SAGE epoch 1 fluxes are connected by dotted lines and epoch 2 by dashed lines.  The sources show SEDs consistent with embedded objects with dust emission.  The sources are marked with solid, red diamonds on the CMDs. The 2MASS data are connected by solid lines and in many instances appears disjoint from either of the epochs, which is due to infrared variability as these data represent a different phase far removed from either of the SAGE epochs. \label{ysosed}}
\end{figure}

In this study,  the variability for this YSO population shows most prominently in the MIPS 24~\micron\ band (24 of 29 YSO candidates have $\vert V_{24}\vert > 3$).   The reason for the predominance of MIPS 24~\micron\ band variability becomes clear when we examine the SEDs  of a few variable YSO candidates (Figure \ref{ysosed}).  The SEDs show the SAGE epoch1 data connected by dotted lines and epoch2 data by dashed lines. The 2MASS data are connected by solid lines and in many instances appears disjoint from either of the epochs, which is due to infrared variability as these data represent a different phase far removed from either of the SAGE epochs.  Most of these sources are very faint in the 2MASS and IRAC bands,  and increase rapidly in flux at the longer wavelengths.  In fact, while our SEDs end at 24~\micron\,  the trend suggests that for some of these YSOs, the SEDs actually peak at even longer wavelengths.   Thus,  our sensitivity to these sources is greatest at 24~\micron\ and the lower number of IRAC variable YSO candidates in the SAGE data may be due to the lower sensitivity in the IRAC bands to detect the variability.     

\citet{whitney08} have completed a candidate list of over 990 YSO candidates identified in the SAGE epoch 1 catalog.   This list is a lower limit to the total number of YSOs in the LMC because the selection was conservative and avoided regions in CMD space densely populated by other types of sources,  e.g. AGB stars and background galaxies.  Our variable YSO candidates comprise $\sim~3\%$ of the larger 990 YSO candidate list.  It is difficult to assess if this percentage is a lower or upper limit because the number of YSOs and variable YSOs is rather uncertain in different directions.  Nevertheless,  the percentage is significant enough to catch our attention as a new class of infrared variables.  With only two epochs of observation spread 3 months apart, we cannot distinguish between a periodic or a more          
stochastic phenomenon for the YSOs.  Clearly, further study of these sources is required to understand the nature of their variability.

\section{Conclusions}
We have detected infrared variable stars in the LMC using the SAGE survey. The variable source population is dominated by evolved stars, in particular the AGB stars (81\% of the SAGE variables). We preferentially find the more evolved, dusty  AGB stars which probably have longer periods.  We also present the discovery of a YSO candidate population with infrared variability. These variable YSO candidates comprise about 3\% of all the SAGE variables. 
\acknowledgments
The SAGE Project is supported by NASA/Spitzer grant 1275598 and NASA NAG5-12595.   We appreciate the support of         
Bernie Shiao,  the STScI database programmer, in the support of this effort.  

{\it Facilities:} \facility{SPITZER (IRAC)}, \facility{SPITZER (MIPS)}.

\end{document}